\documentclass[runningheads]{llncs}
\usepackage{color}
\usepackage{graphicx}
\newcommand{\ie}{\textit{i.e.}, }
\newcommand{\eg}{\textit{e.g.}, }

\begin{document}
\title{Transfer learning to decode brain states reflecting the relationship between cognitive tasks}
\titlerunning{Transfer learning reflects the relationship between cognitive tasks}
%
\author{Youzhi Qu\inst{1}\orcidID{0000-0002-1056-3996} \and
Xinyao Jian\inst{2}\orcidID{0000-0001-6322-9273}\and
Wenxin Che\inst{1}\orcidID{0000-0003-1093-5398}
\and
Penghui Du\inst{1}\orcidID{0000-0002-1422-0501}
\and
Kai Fu\inst{1}\orcidID{0000-0001-8855-9965}
\and
Quanying Liu\inst{1}\thanks{Corresponding author: Q.L (liuqy@sustech.edu.cn)}\orcidID{0000-0002-2501-7656}
}
\authorrunning{Qu et al.}
%
\institute{Shenzhen Key Laboratory of Smart Healthcare Engineering, Department of Biomedical Engineering, Southern University of Science and Technology, Shenzhen, 518055, China \and
Department of Statistics and Data Science, Southern University of Science and Technology, Shenzhen, 518055, China}
\maketitle  
\begin{abstract}
Transfer learning improves the performance of the target task by leveraging the data of a specific source task: the closer the relationship between the source and the target tasks, the greater the performance improvement by transfer learning. In neuroscience, the relationship between cognitive tasks is usually represented by similarity of activated brain regions or neural representation. However, no study has linked transfer learning and neuroscience to reveal the relationship between cognitive tasks. In this study, we propose a transfer learning framework to reflect the relationship between cognitive tasks, and compare the task relations reflected by transfer learning and by the overlaps of brain regions (\eg neurosynth). Our results of transfer learning create \textit{cognitive taskonomy} to reflect the relationship between cognitive tasks which is well in line with the task relations derived from neurosynth. Transfer learning performs better in task decoding with fMRI data if the source and target cognitive tasks activate similar brain regions. Our study uncovers the relationship of multiple cognitive tasks and provides guidance for source task selection in transfer learning for neural decoding based on small-sample data.

\end{abstract}
\section{Introduction}
Transfer learning leverages the knowledge in the source domain data to transfer to the target domain with an assumption that the source and target tasks in the model share some common knowledge~\cite{weiss2016survey,zhuang2020comprehensive}. A model pre-trained with source domain data acquires rich high-order knowledge, which can help to learn the target domain task with only a small amount of target domain data. However, transfer learning is not always beneficial. The performance of transfer learning greatly depends on the relationship between tasks in source and target domains. If 
great distinction exists between the two tasks, transfer learning may negatively affect the learning of target domain, which is also called \textit{negative transfer}. Therefore, it is essential to examine the relationship between tasks in transfer learning. ~\cite{zamir2018taskonomy} proposes a computational model to calculate the task affinity matrix by comparing the performance of transfer learning between tasks in computer vision (\eg such as object recognition, depth estimation and edge detection), which provides guidance on how to select source tasks. This is called \textit{taskonomy}~\cite{zamir2018taskonomy}. Although the taskonomy in computer vision has been intensively studied and utilized in transfer learning, the relationship between cognitive tasks is less clear. There are no studies using transfer learning to explore the relationship between cognitive tasks.

Exploring the neural mechanisms of information process in cognitive tasks is critical to the understanding of the brain. Some tasks may activate overlapping brain regions or induce similar brain activity. Thus, the task relations can be reflected at the neural level. 
For example, the fusiform gyrus is involved in multiple tasks, including recognizing faces and understanding the meaning of written words. The activated brain regions when performing tasks with a closer relation have more overlaps than those with less relation, since the former shares neural circuits in cognitive process~\cite{purves2019neurosciences}. This strategy of our brain evolves to improve multitasking and efficiency~\cite{marois2005capacity}.
From a neuroscience perspective, the relationship between cognitive tasks can be reflected in brain region overlap~\cite{najafi2016overlapping}, or in neural representation similarity~\cite{loffler2005fmri,striem2018neural}.
In the era of deep learning, attention has been paid to the similarities and differences between the brain and artificial intelligence (AI) in processing information~\cite{DiCarlo2012HowDT,klindt2017neural}. Neural representation similarity in artificial neurons has been also applied to reflect the relations of task stimulus, especially visual stimuli~\cite{walker2019inception,Yamins2016UsingGD}. Although some studies have found similarities in neural representations of the brain and AI~\cite{ran2021deep,Yang2019TaskRI}; however, few studies have focused on the commonalities in task performance between the human brain and AI.

\begin{figure}
\includegraphics[width=0.95\textwidth]{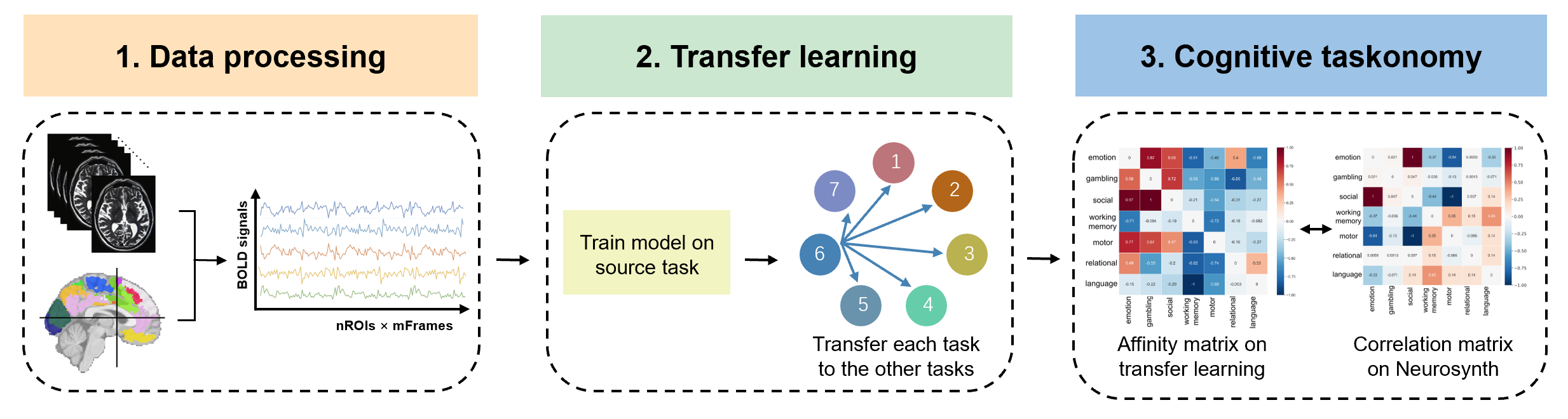}
\caption{Computational modeling of cognitive task relations and creating the \textit{cognitive taskonomy} using transfer learning. From left to right: (1) data preprocessing: preprocessing fMRI data and parcellating brain with atlas; (2) transfer learning: training task-specific networks for task decoding and transferring the network trained with the source task to the rest six tasks; (3) cognitive taskonomy: comparing task affinity matrices derived from transfer learning and Neurosynth.} \label{fig:framework}
\end{figure}

In this study, we present a transfer learning framework to create the relationship between cognitive tasks, called \textit{cognitive taskonomy} (Fig.~\ref{fig:framework}). It obtains a task affinity matrix to represent the relations of various cognitive tasks. The task affinity matrices from transfer learning and from brain activity are compared to examine the resemblance of cognitive taskonomy from 
AI and from the human brain. Our contributions are summarized as follows.
\begin{itemize}
    \item We propose a computational modeling framework of cognitive task relations and create the cognitive taskonomy using transfer learning (Fig.~\ref{fig:framework}).
    \item By comparing the cognitive task affinity matrix derived from transfer learning and from Neurosynth, we uncover a strong resemblance of these two, especially in the emotion, gambling and social tasks (Fig.~\ref{fig:Taskonomy}). %
    \item The affinity of seven cognitive tasks provides guidance for source task selection in transfer learning for brain state decoding (Fig.~\ref{fig:transfer-vs-selfsupervised}\&\ref{fig:transfer-data}).
\end{itemize}

\section{Related Work}
\subsection{Cognitive task relations from neuroscience perspective}
Different cognitive tasks involve a variety of cognitive processes and brain functions. From the neuroscience perspective, the relationship between cognitive tasks can be investigated at different levels, including the overlaps of activated brain regions, the neural representation similarity and the representational similarity in neural decoding models. Similar cognitive processes activate some shared brain regions, so the overlap of activated brain regions can be used to measure the relationship between tasks.
Previous studies have shown that the activation of brain regions exhibits a community structure under cognitive task states, and the overlapping part of the community reflects the functional relationship of cognitive tasks~\cite{najafi2016overlapping,wu2013dissociable}.
Moreover, the similarity of neural representations or the representational similarity of neural decoding models can also reflect the relationship between different cognitive tasks. A large number of neuroscience studies aim to explore the relationship between stimuli and neural responses. For instance, in a vision task, a mapping between facial features and neural representations is constructed~\cite{loffler2005fmri}. Recently, artificial neural networks have been employed as encoding models to construct the mapping from the stimuli to neural responses~\cite{bashivan2019neural,ran2021deep} and as decoding models to classify cognitive states from neural data. Representations of artificial neurons in decoding models are used to measure the similarity of cognitive tasks ~\cite{li2019interpretable,li2021braingnn,Yang2019TaskRI}.

\subsection{Cognitive task relations from transfer learning perspective}

Transfer learning seeks to utilize the knowledge of a source task to a target task. We define a domain in transfer learning as $\mathcal{D}=\{\mathcal{X},P(X)\}$, where $\mathcal{X}$ is the feature space and $P(X)$ is a marginal probability distribution, $X=\{x_1,...,x_n\}$ is sampled from $\mathcal{X}$. In fMRI experiments, $\mathcal{X}$ includes all possible images collected under the same experiment protocol, and $P(X)$ depends on subject groups, such as children or adults. A task is defined as a combination of a label space $\mathcal{Y}$ and a predictive function $f(\cdot)$, i.e. $\mathcal{T}=\{\mathcal{Y},f(\cdot)\}$. The predictive function $f$ is learned from the 
training data $\{(x_{i},y_{i})\}_{i=1}^{n}$, where $x_i\in \mathcal{X}$ and $y_i\in \mathcal{Y}$. Given a source domain $\mathcal{D}_S=\{\mathcal{X}_S,P(X_S)\}$ and task $\mathcal{T}_S=\{\mathcal{Y}_S,f_S(\cdot)\}$, and a target domain $\mathcal{D}_T=\{\mathcal{X}_T,P(X_T)\}$ and task $\mathcal{T}_T=\{\mathcal{Y}_T,f_T(\cdot)\}$, our purpose is to improve the generalization of the target predictive function $f_T$ in $\mathcal{T}_T$ by utilize the knowledge we learned from $\mathcal{D}_S$ and $\mathcal{T}_S$. 

Ideally, we expect that our knowledge acquired from the source domain improves the performance of the target predictive function $f_T(\cdot)$, which is called \textit{transferability}. The transferability is largely decided by the relationship of data in the source and target domains, as well as the source and target tasks. However, in transfer learning, models trained on the some tasks may not be able to be transferred to new tasks. The difference between the source and target domains can have a negative impact when performing transfer learning. For example,  ~\cite{2005To} experimentally shows that if two tasks are too dissimilar, transfer learning may hurt, rather than improve, the performance of the target task.

There have been some studies using transfer learning to evaluate the relationship between transferability and similarity of source and target tasks. For example, ~\cite{zamir2018taskonomy} investigated the underlying relational structure of different tasks by computing the affinity matrix based on the ability to solve one task using representations trained for another task. Their results suggest that the higher similarity of representations among tasks leads to better transferability in transfer learning. ~\cite{achille2019task2vec} developed a TASK2Vec method that can provide a fixed-dimensional embedding of the task. They demonstrated that this embedding can predict task similarities that match our intuition about semantic and taxonomic relations between different tasks, and is also helpful for us to choose a pre-trained model to solve a new task. \cite{dwivedi2019representation}  used representation similarity analysis to obtain a similarity score among tasks by computing correlations between models trained on different tasks. Their results reveal that the higher the similarity score between tasks, the better the transfer learning performance between them.

\begin{figure}
  \centering
  \includegraphics[width=0.95\textwidth]{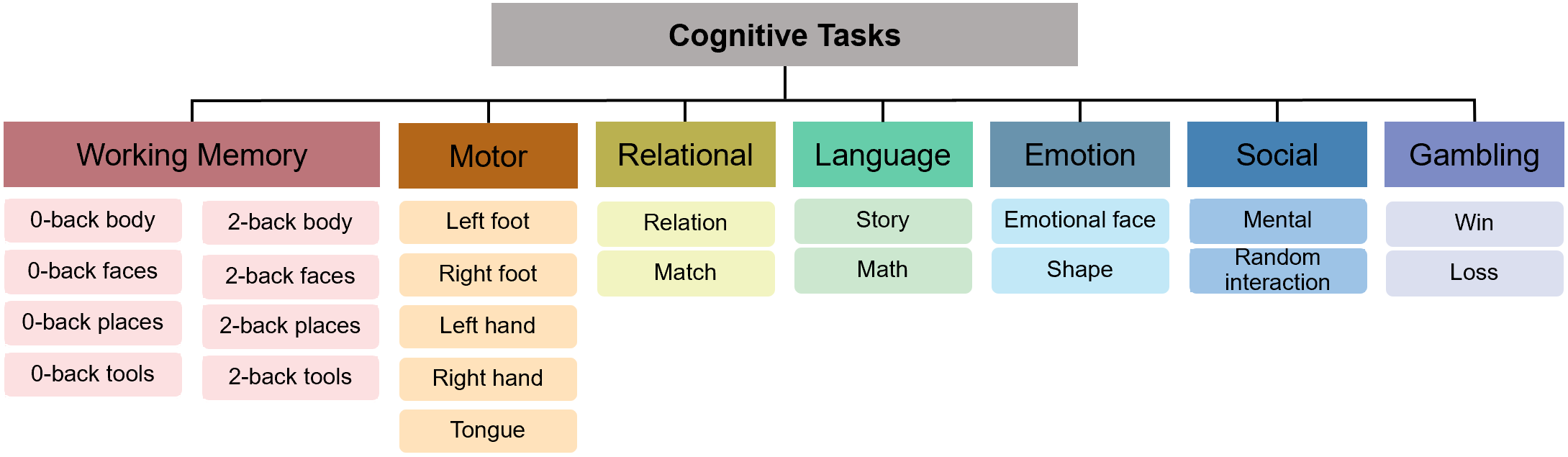}
  \caption{Task Description. HCP fMRI dataset includes 23 tasks, belonging to 7 categories. }
  \label{fig-cog_task}
\end{figure}

\section{Methods}

\subsection{HCP tasks}

Our experiments are conducted on a large fMRI dataset, \ie Human Connectome Project (HCP) S1200 Release~\cite{van2013wu}. The HCP fMRI data are recorded from over 1,000 subjects while they are performing 7 categories of cognitive tasks, including the emotion processing tasks ~\cite{hariri2006preference}, the gambling tasks ~\cite{delgado2000tracking}, the social cognition tasks ~\cite{castelli2000movement}, the working memory tasks ~\cite{drobyshevsky2006rapid}, the motor tasks ~\cite{buckner2011organization}, the relational processing tasks ~\cite{smith2007localizing} and language processing tasks ~\cite{binder2011mapping}. Each category consists of 2-8 subtasks (Fig.~\ref{fig-cog_task}). To eliminate the effect of different number of subtasks on transfer learning, we choose 0-back faces and 2-back faces as subtasks of working memory category, and left hand and right hand as subtasks of motor category. The HCP fMRI data are preprocessed using the standard HCP pipeline for removing spatial distortions, motion correction, registration and normalization. We then parcellate the brain into 90 regions using automatic anatomical labeling (AAL) atlas ~\cite{tzourio2002automated}. The preprocessed HCP task-based fMRI data is detailed in Table~\ref{table:hcptask}.

\begin{figure}
    \centering
    \includegraphics[width=0.92\textwidth]{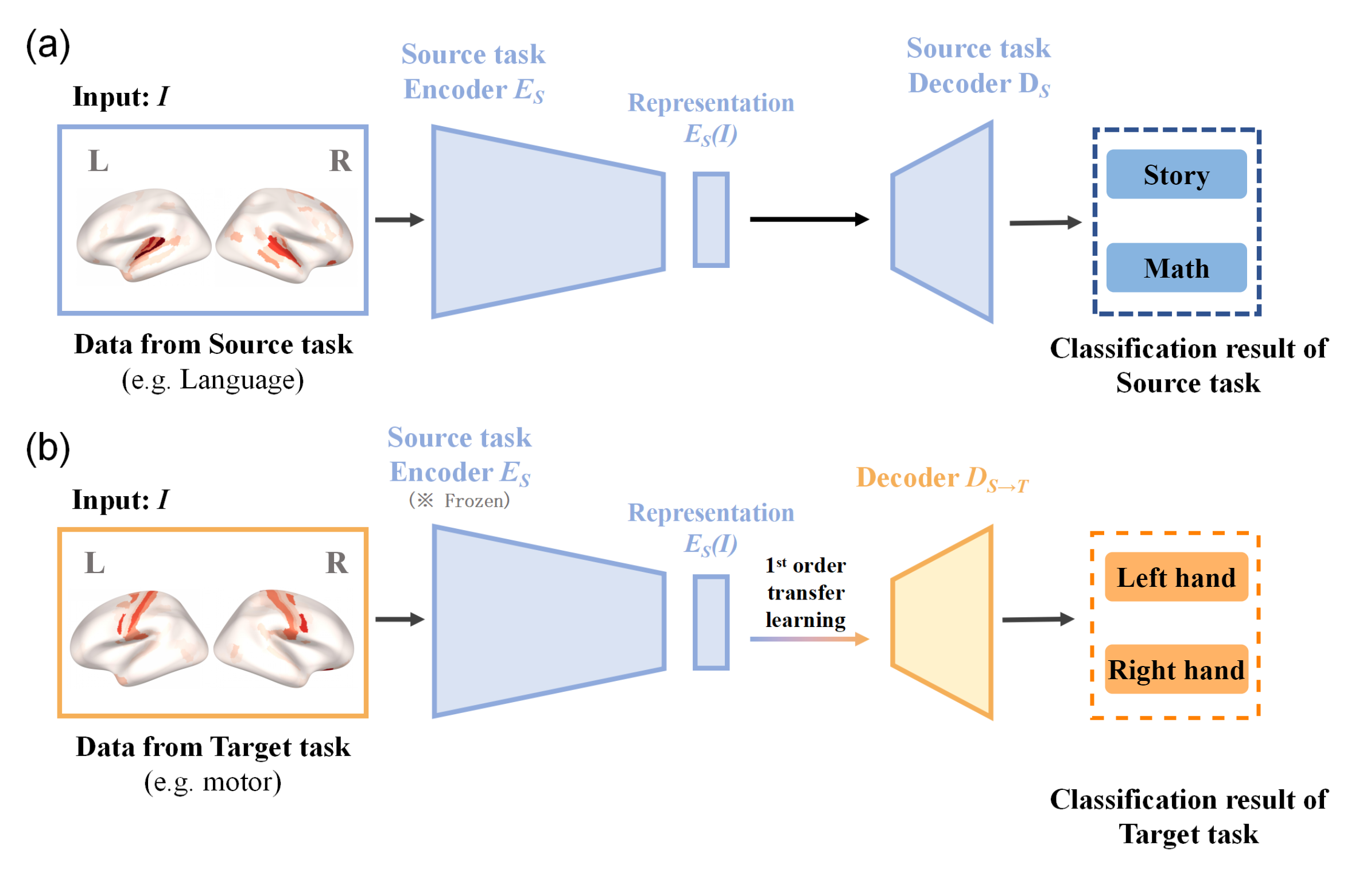}
    \caption{Using transfer learning to estimate the affinity between the source task and the target task. (a) We first train 7 task-specific networks for classifying the tasks in each category with fMRI data as input in a supervised manner. (b) We use the fMRI data from the target task as the input $I$ of the encoder $E_S$ trained on the source task $S$. The parameters of the encoder are frozen during transfer learning. We train a decoder $D_{S\rightarrow T}$ with the representation $E_S(I)$ in the encoder as the input to perform classification on the target task $T$. The classification accuracy from 7 source tasks to 7 target tasks is the $7\times 7$ task affinity matrix.}
    \label{fig:training}
\end{figure}

\begin{table}[ht]
\caption{Details of HCP task-based fMRI data in our study.}
\label{table:hcptask}
\begin{center}
\begin{tabular}{lcccc}
\hline
Category of tasks & $\#$ of subjects & $\#$ of subtasks for classification\\
\hline
    Working memory & 1077 & 2\\
    Motor& 1076 & 2\\
    Relational processing & 1036 & 2\\
    Language processing & 1040& 2\\
    Emotion processing & 1040 & 2\\
    Social cognition & 1044 & 2\\
    Gambling & 1080 & 2\\
\hline
\end{tabular}
\end{center}
\end{table}

\subsection{Transfer Learning}
For each category of tasks, we first train a 4-layer fully connected neural network with fMRI data as the input to perform classification among subtasks. Then, we transfer the seven task-specific pre-trained neural networks to the target domain task, resulting in a \textit{task affinity matrix} to reflect the relations of cognitive tasks (Fig.~\ref{fig:training}).

\textbf{Network Architecture}: There are a total of seven task-specific networks for classification tasks, which correspond to 7 cognitive categories. For each classification task, the classification model should identify the specific task states under the category. The corresponding relationship between categories and tasks is shown in Fig.~\ref{fig-cog_task}. The classification model is divided into two parts: encoder and decoder. For each task, we use the same model architecture consistent to avoid additional bias. Specifically, both the encoder and decoder consist of 2 fully connected layers. 

\textbf{Training task-specific networks}: We first train 7 task-specific networks for classifying the tasks in each category with fMRI data as input in a supervised manner. These 7 task-specific networks are treated as the gold standard of source domain. Specifically, the task-related fMRI data in the source domain is divided into training set, validation set and test set with the ratio of 8:1:1. All models are trained using the Adam optimizer with an initial learning rate of 0.001 and cross-entropy loss. The maximum epoch number is set to 50. When model performance stopped improving on the validation set at 5 epochs, we apply the early stopping criterion to stop training. The performance of decoding cognitive tasks is quantified with classification accuracy.

\textbf{Transfer Learning}: After the gold-standard task-specific neural networks are trained, we perform transfer learning. We treat each task-specific network for classifying one of the 7 categories as the source task, and the rest 6 categories as the target task separately. In transfer learning, we fix the parameters of the two fully connected layers of the encoder in the model, and train the model with 1\%, 5\%, 10\%, 20\%, 30\%, 40\% and 50\% proportions of target domain data as training set, respectively. Another 10\% proportion of target domain data is used as the validation set, and early stopping is applied to reduce overfitting. For each proportion of the transfer results, we repeat 10 times and average the accuracy.

\subsection{Validation of cognitive taskonomy} 
To validate the cognitive taskonomy obtained by transfer learning, we used the task-related brain regions from Neurosynth~\cite{yarkoni2011large}, an fMRI meta-analysis platform. Specifically, we input 7 keywords (\ie working memory, motor, relational, language, emotions, social cognition, gambling) to Neurosynth for retrieving the brain topological maps corresponding to the 7 categories of tasks. Then, we compute the Pearson correlation of the brain maps, representing the similarity matrix across tasks. 
We set diagonal of the similarity matrix to 0, and normalize positive and negative values in the matrix to the intervals 0 to 1 and from 0 to -1, respectively.

\begin{figure}
    \centering
    \includegraphics[width=\textwidth]{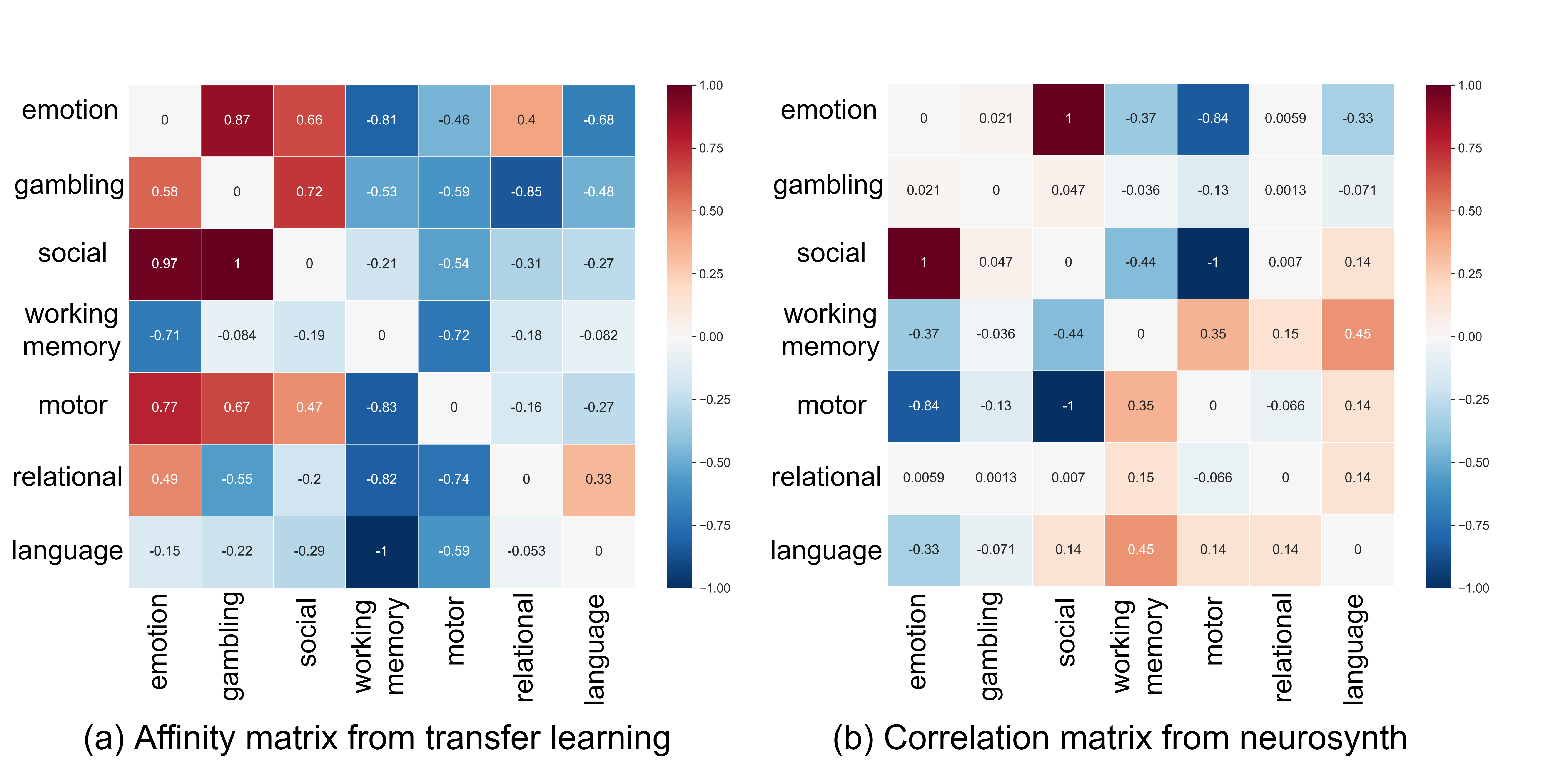}
    \caption{Cognitive taskonomy. (a) The accuracy change from the source task to the target task using transfer learning. The positive value represents that the source task contributes to improving the accuracy of target task, indicating the similarity between the source and task tasks; vice versa. (b) The correlation of cognitive tasks derived from the brain regions. The task-related brain regions are exported from Neurosynth, an online fMRI meta-analysis platform.}
    \label{fig:Taskonomy} 
\end{figure}

\section{Results}
\subsection{Affinity matrix of cognitive tasks}
We use the performance of training task-specific supervised networks with 1\% target domain data as the baseline, and check whether using different source domain data for transfer learning can improve the performance of brain decoding. Fig.~\ref{fig:Taskonomy} (a) shows the accuracy changed from the source task to the target task between seven categories. The row in figure represents the source task, the column in figure represents the target task. We normalize positive and negative values in the matrix to the intervals 0 to 1 and from 0 to -1, respectively, which is convenient for comparing the relationship between different source tasks and the target tasks.
The $(i,j)$-th element of the affinity matrix in Table~\ref{fig:Taskonomy}(a) represents the transferring from the $i$-th task to the $j$-th task. For instance, the first three categories of tasks (\ie emotion, gambling, and social tasks) show positive transfer to each other. In contrast, they have negative transfer to working memory and motor tasks. 
Fig.~\ref{fig:Taskonomy} (b) show the Pearson correlation coefficient of the brain topological maps from Neurosynth. The results show that this affinity matrix is high resemble of the task affinity matrix from transfer learning, especially the emotion, gambling and social tasks. Moreover, the tasks with more overlaps in brain regions (\eg emotion, gambling and social) can better improve the brain decoding performance of the target tasks in transfer learning, while the negative correlation in activity maps brings negative transfer (\eg emotion to working memory, emotion to motor).

\begin{figure}
    \centering
    \includegraphics[width=.7\textwidth]{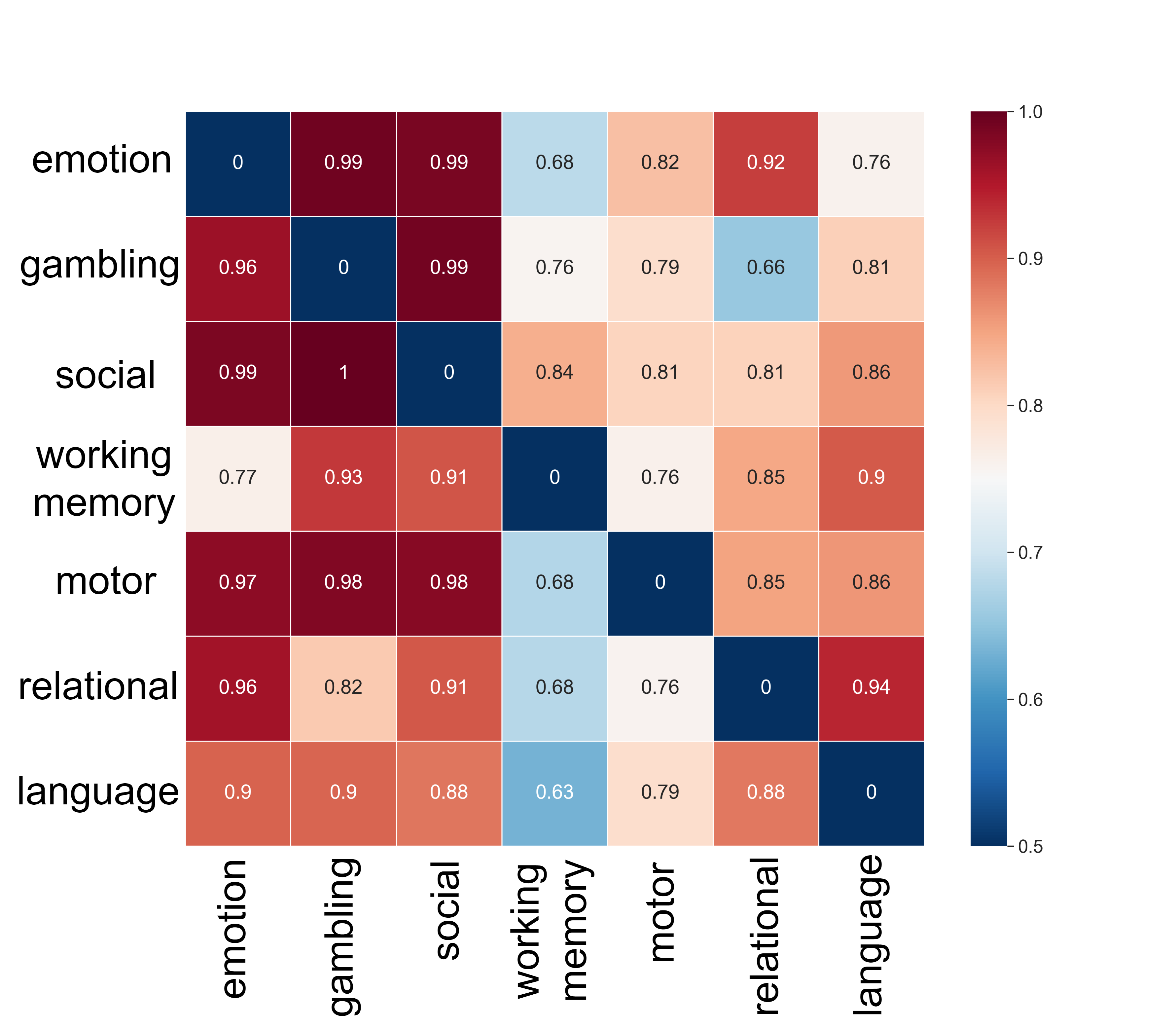}
    \caption{The accuracy comparison of transfer learning with 1\% data and supervised learning with 80\% data. The value represents the ratio of transfer learning performance to supervised learning performance. The larger the value, the closer the performance of transfer learning is to the gold standard.}
    \label{fig:transfer-vs-selfsupervised}
\end{figure}

\subsection{Compare with task-specific networks}
We use the performance of training task-specific supervised networks with 80\% target domain data as the gold standard. We compare the brain decoding performance of the gold standard with transfer learning using 1\% target domain data. Fig.~\ref{fig:transfer-vs-selfsupervised} shows the ratio of the decoding performance of transfer learning to the gold standard. The larger the ratio, the closer the transfer performance is to the task-specific network performance. The result shows that when the three categories, emotion, gambling and social are used as the target tasks, the transfer learning from source task for other categories can achieve the performance close to the gold standard when using a small amount of data. However, when the working memory is used as target tasks, the effect of transfer learning is not good.

\begin{figure}
    \centering
    \includegraphics[width=.95\textwidth]{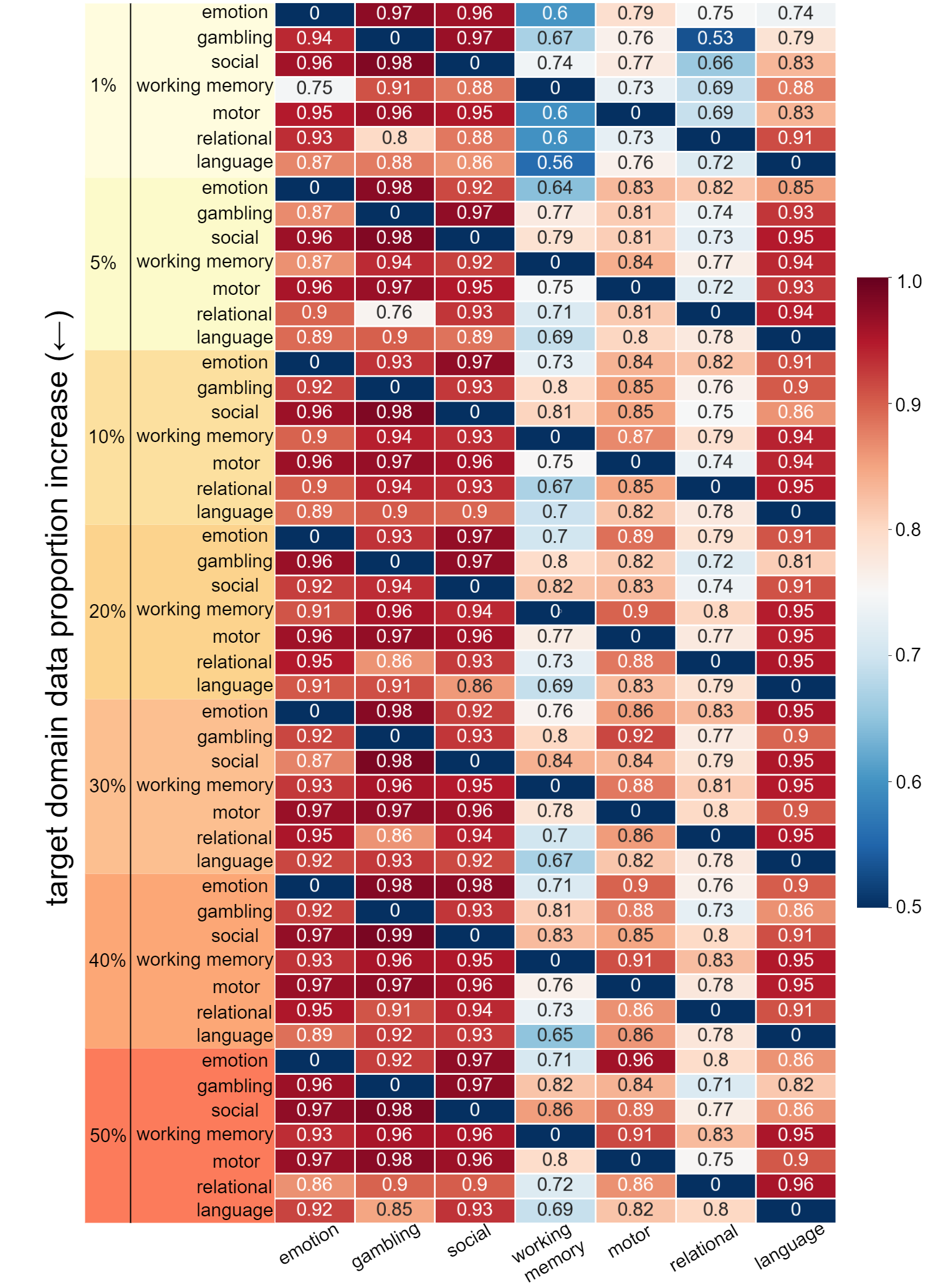}
    \caption{The performance in transfer learning. The model is trained using 1\%, 5\%, 10\%, 20\%, 30\%, 40\% and 50\% of target domain data, respectively. The category in the row represents the source task and the category in the column represents the target task. The values represent the accuracy of the transfer learning model on target tasks in test sets.}
    \label{fig:transfer-data}
\end{figure}

\subsection{Brain decoding accuracy with transfer learning}
Fig.~\ref{fig:transfer-data} shows the performance of transfer learning between 7 categories when using various proportions of target domain data ranging from 1\% to 50\%. The brain decoding accuracy of the target task increases with the amount of target domain data. No matter what the source task is, transfer learning on three target domain tasks (\ie emotion, gambling and social) is consistently higher than other target tasks. Remarkably, the classification accuracy achieves over 90\% in these four tasks with only 1\% of target domain data. In contrast, the classification accuracy is much lower in working memory category. 
These results indicate that the cognitive taskonomy is informative for transfer learning.

\section{Discussion} 

\textbf{Significance in neuroscience}:
Cognitive taskonomy reveals the relationship of cognitive processes in the brain at cognitive tasks. Our results show the task affinity matrix from learning is similar to the correlation matrix from Neurosynth  (Fig.~\ref{fig:Taskonomy}). For instance, transfer learning between emotion, gambling and social categories achieve high performance, indicating a close relationship between these tasks. This finding is in line with the neuroscience study which reports that these cognitive tasks share some cognitive process in our brain~\cite{britton2006neural,li2010iowa}. In contrast, performing transfer learning from these three categories to the other categories (\eg motor), the accuracy of the target task is not improved, suggesting a distinct cognitive process between tasks. It is consistent with the brain maps exported from Neurosynth~\cite{yarkoni2011large}, and the prior knowledge of brain networks~\cite{liu2017detecting}. This resemblance between the task relationships obtained from transfer learning's performance and the task relationships obtained from brain regions implies that AI and the brain are coherent in processing cognitive tasks.

\textbf{Significance in transfer learning}: Cognitive taskonomy is informative for the selection of source tasks in transfer learning for neural decoding applications, especially when the training data is extremely small. Transfer learning performance between 7 cognitive categories (Fig.~\ref{fig:transfer-vs-selfsupervised}) indicates that when the emotion, gambling and social cognitive categories are used as source tasks to perform transfer learning on other categories, the accuracy of the target tasks is close to the gold standard, even using only 1\% target domain data. However, when the working memory cognitive category is used as source task, the accuracy of the target tasks is poor. The results indicate that the emotion, gambling and social cognitive categories are transferable, while the working memory~\cite{wang2020decoding} cognitive category is non-transferable. Fig.~\ref{fig:transfer-data} shows that the transferability and non-transferability of cognitive tasks are robust, and they do not change when using different proportions of target domain data. This non-transferability may be related to decoding uncertainty. For instance, researchers have shown that working memory is not processed in a single brain site, but stored and processed in widely distributed brain regions~\cite{christophel2017distributed}. The distributed nature
of the working memory cognitive task leads to a hard problem to decode perfectly.

\textbf{Limitations and future work}: In this study, we used a transfer learning model based on fully connected layers which are relatively simple. The correlations between cognitive tasks may not be fully explored.
The experiment is based on the classification performances to evaluate the transfer performances. 
Performing more complex decoding tasks on cognitive tasks, such as predicting fMRI signals, is a better way to explore the relationship between cognitive tasks. Whether the similarity between neural representations in the model, as well as the similarity in transfer learning performance changes between cognitive tasks, has the ability to describe the relationship between cognitive tasks, these are worth exploring in the future.

\textbf{Conclusion}: We propose a transfer learning framework to create cognitive taskonomy. The results demonstrate the similarities between brain intelligence and artificial intelligence. Furthermore, cognitive taskonomy opens a new window for source task selection in transfer learning for brain state decoding using small data.

\section*{Acknowledgments}
This work was funded in part by the National Key Research and Development Program of China (2021YFF1200804), National Natural Science Foundation of China (62001205), Guangdong Natural Science Foundation Joint Fund (2019A1515111038), Shenzhen Science and Technology Innovation Committee (20200925155957004, KCXFZ2020122117340001), Shenzhen-Hong Kong-Macao Science and Technology Innovation Project (SGDX2020110309280100), Shenzhen Key Laboratory of Smart Healthcare Engineering (ZDSYS20200811144003009).

%
%
%
\bibliographystyle{splncs04}
\bibliography{reference}
%




\end{document}